\documentclass[preprintnumbers,eqsecnum,amsmath,amssymb,nofootinbib]{revtex4}
\usepackage{graphicx}
\usepackage{dcolumn}
\usepackage{bm}
\usepackage{epsfig}

\begin{document}
\title{Width of the $\bm{ J^P={\frac{1}{2}}^+}$ pentaquark in the quark-diquark model} 
\author{D.~Melikhov$^{a,b}$, S.~Simula$^c$, and B.~Stech$^{a}$}
\affiliation{
$^a$ Institut f\"ur Theoretische Physik, 
Universit\"at Heidelberg, 
Philosophenweg 16, 69120, Heidelberg, Germany\\
$^b$ Nuclear Physics Institute, Moscow State University, 119991, Moscow, Russia\\
$^c$ INFN, Sezione di Roma Tre, Via della Vasca Navale 84, 00146, Roma, Italy}
\begin{abstract}
We analyse the width of the $\theta(\frac12^+)$ pentaquark assuming that it is a bound state
of two extended spin-zero $ud$-diquarks and the $\bar s$ antiquark (the Jaffe-Wilczek scenario).
The width obtained when the size parameters of the pentaquark wave function are taken close to
the parameters of the nucleon is found to be $\simeq 150$ MeV, i.e. it has a normal value for
a $P$-wave hadron decay with the corresponding energy release.
However, we found a strong dynamical suppression of the decay width 
if the pentaquark has an asymmetric "peanut" structure with the strange
antiquark in the center and the two diquarks rotating around. In this case a decay width 
of $\simeq$ 1 MeV is a natural possibility. 
\end{abstract}
\maketitle
\section{Introduction}
A narrow hadron of the suspected quark content $udud\bar s$,
the Pentaquark $\theta(1540)$, has been reported by several experiments \cite{signal}. 
In addition, a narrow charmed exotic state with the presumed content $udud\bar c$ has been
observed \cite{hera}. Nevertheless, there still remain doubts because of the negative
results from some other experiments \cite{nosignal} and problems with the identification of 
pentaquark states with different flavours \cite{nussinov,glozman}. 
A striking feature of the pentaquark is its width of only several MeV \cite{trilling}. 
The pentaquark has most likely spin 1/2 and isospin 0. Its parity is not yet known.

Several interpretations of this state have been discussed in the literature 
(see \cite{nussinov,close} and refs therein).
Most likely it is the narrow exotic state predicted by Diakonov, Petrov and Polyakov \cite{soliton} 
using a chiral soliton model. For a discussion of the width obtained in this model 
and of earlier suggestions \cite{mp} we refer to \cite{jaffe}.

Our goal is to study the width of the new state in a quark model. To this end we follow the
suggestion of Jaffe and Wilczek \cite{jw} that the correlation between two quarks in a $0^+$ color
antitriplet state is of particular importance. Their
pentaquark model consists therefore of two scalar isospin
zero diquarks in a relative angular momentum $L=1$ state and the strange antiquark. The
parity of this state is even. In our view this configuration is very likely since the
correlation between two quarks in a
color antitriplet state forming scalar diquarks is as strong as the corresponding $\bar qq$ 
correlation in a $\pi$ meson. This is known from QCD sum rules
\cite{stech} and from instanton calculations \cite{shuryak}. The most dramatic effect of
diquark formation manifests itself in strangeness changing weak
decays at low energies: It causes the huge $\Delta I = 1/2$ enhancement
present in these processes \cite{huge}. 

The $\theta\to NK$ decay amplitude can be obtained from the single constant $g_A$, the
$\theta\to N$ axial vector form factor at $q^2=m_K^2$.
In the analogous nucleon-to-nucleon transition one has $g_A^N(q^2=0)\simeq 1.25$. For a $P$-wave
resonance with a mass of 1540 MeV the value $g_A=0.8$ would lead to the
width $\Gamma(\theta)=\Gamma(\theta\to p K^0)+\Gamma(\theta\to n K^+)= 160$ MeV.
To have $\Gamma(\theta)\le 10$ MeV one needs a strong suppression $g_A\le 0.2$.
Several recent papers have discussed possible reasons for a suppression of the 
$\theta\to NK$ transition amplitude by describing the spin-flavour and color factors in
the $\langle NK|\theta\rangle $ overlap amplitude \cite{close,lipkin,zahed,carlson,buccella}. 
But a complete model calculation has so far not been performed. 

We calculate the width of the pentaquark in a non-relativistic Fock space description
using three and five constituent quarks for the proton and the pentaquark, respectively. 
A non-relativistic treatment seems
appropriate since the spatial momentum of the final nucleon (in the rest
system of the pentaquark) is not large. We use the quark-diquark picture for the
proton and the pentaquark taking the size of the diquarks similar to the pion size. 
This appears to be justified from the near equality of the
decay constants of pion and diquark and from the agreement of the proton
decay constant using a spatially extended diquark with other determinations \cite{stech,huge}. 
The essential ingredients in our calculation are the size parameters of the pentaquark
wave function for the diquark-diquark-antiquark configuration.
If all size parameters of the $\frac{1}{2}^+$ pentaquark wave function are taken to
be close to the size parameters of the nucleon, we obtain the width $\Gamma(\theta)\simeq 150$ MeV.
On the other hand, the width is found to be very strongly suppressed when 
the pentaquark has a significantly asymmetric structure ("peanut"-like), in which the strange
antiquark remains near to the center and the two diquarks are rotating around it. A width of $\simeq$
1 MeV is then easily obtainable.

We can also conclude from our investigation  that any strong suppression of the transition
amplitude must have a dynamical origin in the spatial part of the pentaquark wave function. 
We did not find a sizeable suppression of the decay amplitude due to the color-flavour structure of the 
pentaquark.

The paper is organized as follows: 
In the next section we discuss the relationship between the $\theta\to NK$ amplitude and the
form factor $g_A$ obtained from the axial-vector matrix element
$\langle N|\bar s\gamma_\mu\gamma_5 d|\theta\rangle$.
In section 3 we describe the nucleon and the pentaquark in terms of quark creation operators
and calculate $g_A$ in terms of the corresponding wave functions according to the quark diquark
picture. Section 4 contains the numerical study of $g_A$ and the width for various choices of  
the parameters for the pentaquark wave function. Section 5 presents our conclusions.

\section{The $\theta\to KN$ decay and the form factor $g_A$}
The formula for the two-body decay width of the $\theta$ resonance has the form 
\begin{eqnarray}
\label{width1}
\Gamma(\theta\to KN)=\frac{1}{8\pi}
\frac{|\vec q|}{M_\theta^2}\sum_{\rm final\; spins}{|T(\theta\to KN)|^2}, 
\end{eqnarray}
where the sum runs over nucleon polarizations  and $|\,\vec q\,|$ denotes the momentum of the 
outgoing nucleon in the $\theta$ rest frame
\begin{eqnarray}
|\,\vec q\,|=\frac{1}{2M_\theta}{\sqrt{(M_\theta^2-M_N^2-M_K^2)^2-4 M_N^2 M_K^2}}. 
\end{eqnarray}
The decay amplitude $T(\theta\to KN)$ is related to the $S$-matrix according to  
\begin{eqnarray}
\langle  N(p')K(q)|S|\theta(p\rangle=
i(2\pi)^4\delta(p-p'-q)T(\theta\to NK) 
\end{eqnarray}
for the standard relativistic normalization of states  
$\langle \vec p|\vec p\,'\rangle=2p^0(2\pi)^3\delta(\vec p-\vec p\,')$.  
The amplitude may be expressed by the 
$\theta\to N$ matrix element of an interpolating field $\phi_K$ of the kaon 
\begin{eqnarray}
T(\theta\to NK)=\lim_{q^2\to M_K^2}(M_K^2-q^2)\langle N|\phi_K(0)|\theta\rangle. 
\end{eqnarray}
Without loss of generality, $\phi_K$ can be taken to be the divergence of the strangeness-changing axial-vector 
current $A_\mu$:  
\begin{eqnarray}
A_\mu=\bar s^a\gamma_\mu\gamma_5 d^a, \qquad 
\phi_K=\frac{1}{f_KM_K^2}\partial^\mu A_\mu, 
\end{eqnarray}
where $f_K$ denotes the kaon decay constant ($f_K=0.160$ GeV). 
The matrix element of the axial current between the $\theta$ and the nucleon can be decomposed 
in terms of invariant form factors
\begin{eqnarray}
\label{2.7}
\langle N(p')|\bar s\gamma_\mu \gamma_5 d|\theta(p)\rangle = 
g_A(q^2) \bar u_N(p')\gamma_\mu\gamma_5 u_\theta(p)
+
g_P(q^2) q_\mu \bar u_N(p')\gamma_5 u_\theta(p)
+
g_T(q^2) \bar u_N(p')\sigma_{\mu\nu}q^\nu \gamma_5 u_\theta(p)
\end{eqnarray}
with $q=p-p'$. The spinor amplitudes for both $\theta$ and $N$ are normalized according to relation $\bar u_i(p) u_i(p)=2M_i$, 
$i=N,\theta$. The form factors $g_i$ contain poles at $q^2>0$ due to strange meson resonances 
with the appropriate quantum numbers: e.g. $g_P$ contains the $K$-meson pole and  
$g_A$ the $K^*_A$ pole. 

In the chiral limit $m_s=m_d=M_K=0$, the axial current is conserved $\partial_\mu A^\mu=0$ and the form factors 
$g_A$ and $g_P$ are related to each other. The form factor $g_P(q^2)$ has a pole at $q^2=0$ 
corresponding to the massless kaon  
\begin{eqnarray}
\label{chiral}
g_P(q^2)=\frac{M_\theta+M_N}{q^2} g_A(q^2). 
\end{eqnarray}
For $m_s\ne 0$ and $M_K\ne 0$, the pole in the form factor $g_P$ is shifted to $M_K^2$, and 
the form factors $g_A$ and $g_P$ are no longer related to each other by (\ref{chiral}). However, 
due to the smallness of the strange-quark mass the main effect of its nonzero value 
results in the shift of the location of the pole in $q^2$. We then have the approximate relation 
\begin{eqnarray}
g_P(q^2)=-\frac{M_\theta+M_N}{M_K^2-q^2} g_A(q^2), 
\end{eqnarray}
and in particular the PCAC result  
\begin{eqnarray}
\label{2.8}
\langle N(p')|\partial_\mu A^\mu|\theta(p)\rangle=
\bar u_N(p')i\gamma_5 u_\theta(p)(M_\theta+M_N)g_A(q^2) \frac{M_K^2}{M_K^2-q^2}. 
\end{eqnarray}
Therefore, the decay amplitude is known if we know $g_A=g_A(M_K^2)$: 
\begin{eqnarray}
\label{2.11}
T(\theta\to NK)=\frac{M_\theta+M_N}{f_K}g_A\cdot \bar u_N(p')i\gamma_5 u_\theta(p).  
\end{eqnarray}
The decay width (\ref{width1}) takes the simple form 
\begin{eqnarray}
\label{2.12}
\Gamma(\theta\to K^+n)=\Gamma(\theta\to K^0p) 
=\frac{1}{2\pi}\frac{|\vec q|^3}{f_K^2}\frac{(M_\theta +M_N)^2}{(M_\theta +M_N)^2-M_K^2}g_A^2\simeq  
\frac{1}{2\pi}\frac{|\vec q|^3}{f_K^2}g_A^2. 
\end{eqnarray}
For $M_\theta=1540$ MeV one finds $|\vec q|=270$ MeV. The corresponding width is 
$\Gamma(\theta)=\Gamma(\theta\to K^+n)+\Gamma(\theta\to K^0p) \simeq  240\; g_A^2$ MeV. 
Typically, for transitions between hadrons of the same quark structure one has $g_A\simeq 1$ 
(e.g. for the nucleon $g_A\simeq 1.23$), such that for a normal resonance with  mass $M=1540$ MeV 
one would expect $\Gamma(\theta(1540)\to KN)\simeq 120\div 180$ MeV \cite{jaffe}.  
To obtain a width of $\le$ 10 MeV one needs a strongly suppressed value $g_A\le 0.2$. 

We shall obtain the form factor $g_A$ by calculating different components of the l.h.s.~of 
(\ref{2.7}) for different polarizations of the initial $\theta$, working in 
the $\theta$ rest frame $p=(M_\theta, \vec{0})$ and 
choosing $q=(q_0,0,0,|\vec q|)$. It is convenient to use now the following normalization of states 
\begin{eqnarray}
\label{nrnorma}
\langle \vec p|\vec p\,'\rangle=(2\pi)^3\delta(\vec p-\vec p\,').  
\end{eqnarray}
Then $g_A$ can be obtained from the equation 
\begin{eqnarray}
\label{ga}
g_A(q^2)&=&(M_N-M_\theta)\frac{(M_\theta+M_N)^2-q^2}{4 M^2_\theta} F_0 
+\frac{M_N+M_\theta}{2M_\theta}A_L -\frac{q^2(A_T-A_L)}
{(M_N-M_\theta)^2-q^2}
\end{eqnarray}
with
\begin{eqnarray}
\label{components}
F_0=\frac{1}{|\vec q|}\langle N^\uparrow(\vec p')| \bar s\gamma_0\gamma_5 d|\theta^\uparrow\rangle, \qquad   
A_L=\langle N^\uparrow(\vec p')| \bar s\gamma_3\gamma_5 d|\theta^\uparrow\rangle, \qquad 
A_T=i\langle N^\uparrow(\vec p')|\bar s\gamma_2\gamma_5 d|\theta^\downarrow\rangle.  
\end{eqnarray}
$\vec p'=-\vec q$ lies in the negative $z$-direction, $|\vec q|=\sqrt{(M_\theta^2-M_N^2-q^2)^2-4 M_N^2 q^2}/(2M_\theta)$.
Note that $A_L=A_T$ for $\vec q=0$. 

In a model calculation of the left-hand side of (\ref{2.7})
one cannot expect to obtain the $K$-meson pole. This means that not all the 
components of the current, and therefore not all form factors  
can be calculated in a quark model for the nucleon and the pentaquark. 
However one can expect that the linear combination of the current components relevant for the form factor 
$g_A$, which does not have a pole in the physical $\theta\to N$ decay region 
(the $K^*_A$ resonance which leads to the pole in $g_A$ lies at a much higher scale), 
can be calculated reliably within the quark model.


\section{The $\theta\to pK^0$ amplitude from a nonrelativistic quark model}

In this section we calculate the amplitudes 
$\langle N(-\vec q)|\bar s \gamma_\mu \gamma_5 d|\theta\rangle$ 
and the form factor $g_A(q^2)$ using a non-relativistic equal-time Fock space representation.  
To do so we replace the strange quark field $s$ by its charge conjugate field 
$s_c=C(\bar s)^T$ which annihilates the strange antiquark in $\theta$. 
$C=i\gamma_2\gamma_0$ is the charge conjugation matrix. 

To make notations simpler we imply that the operator $\hat s$ creates the $\bar s$-quark from the vacuum, 
and the operators $u$ and $d$ create $u$ and $d$ quarks, respectively. 

Let us describe now the states involved in the decay, specifically for the process $\theta\to pK^0$. 

\subsection{The nucleon}
In terms of three constituent quarks the proton state can be 
written in the form \cite{stech}
\begin{eqnarray}
\label{nucleon}
|N^{\uparrow}(\vec p)\rangle&=& 
\int d\vec r_1 d\vec r_2 d\vec r_3
e^{i\vec p\,\vec R}\;\Psi_N(\vec r_1|\vec r_2,\vec r_3)\;
u^{a\uparrow}(\vec r_1)D^a(\vec r_2, \vec r_3)|\,0\rangle, 
\end{eqnarray}
where we have introduced a bilocal creation operator $D$ for two quarks in a color-antitriplet state: 
\begin{eqnarray}
D^a(\vec r_2, \vec r_3)&=&\epsilon^{aa_2a_3}
\left[u^{a_2\uparrow}(\vec r_2)d^{a_3\downarrow}(\vec r_3)
-
u^{a_2\downarrow}(\vec r_2)d^{a_3\uparrow}(\vec r_3)\right]  
\end{eqnarray}
and 
\begin{eqnarray}
\label{3.6}
\vec R=\frac{1}{3}(\vec r_1+\vec r_2+\vec r_3). 
\end{eqnarray}
The radial wave function $\Psi_N(r_1|r_2,r_3)$ must be symmetric under $2\leftrightarrow 3$ to 
guarantee that the nucleon has spin and isospin 1/2. 
It depends on the relative coordinates 
$\rho$ and $\lambda$ 
\begin{eqnarray}
\label{3.6b}
\vec\rho=\vec r_2-\vec r_3,\qquad 
\vec\lambda=\frac1{2}(\vec r_2+\vec r_3)-\vec r_1. 
\end{eqnarray}
i.e. 
\begin{eqnarray}
\label{psin}
\Psi_N(r_1|r_2,r_3)=\Psi_N(\vec{\rho},\vec{\lambda}).
\end{eqnarray}
The state normalization (\ref{nrnorma}) leads to the following 
condition for the proton radial wave function \cite{stech}
\begin{eqnarray}
\label{nucleonnorma}
6\int d\vec\rho \, d\vec \lambda \, \Psi_N\left(\vec\rho,\vec\lambda\right)
\left[
2\Psi_N\left(\vec\rho,\vec\lambda\right)+
\Psi_N\left(\frac12\vec\rho-\vec\lambda,-\frac34\vec\rho-\frac12\vec\lambda\right)\right]=1. 
\end{eqnarray}
We parametrize the wave function by harmonic oscillator functions in $\rho^2$ and $\lambda^2$ which correspond 
to assuming harmonic forces between the constituents:  
\begin{eqnarray}
\label{nucleonwf}
\Psi_N(\rho,\lambda)=\frac{1}{\sqrt{N_p}}\exp\left(-\frac{1}{2\alpha^2_{\rho N}}\vec\rho^2 
-\frac{2}{3\alpha^2_{\lambda N}}\vec\lambda^2\right).  
\end{eqnarray}
The normalization factor $N_p$ is determined by (\ref{nucleonnorma}).  
A fully symmetric function under coordinate exchanges corresponds to 
$\alpha_{\rho N}=\alpha_{\lambda N}$.\footnote{We emphasize that for the symmetric case 
the nucleon wave function (\ref{nucleonwf}) is fully equivalent to the standard SU(6) 
symmetric 56-plet wave function. By reordering it describes    
along with the spinless $ud$ diquark made of quarks 2 and 3, the spin-1 
$ud$ "diquark" made of particles 1 and 3. The quark-diquark form of the nucleon wave function 
becomes more than just a notation if the correlation between the $u$ and $d$ quarks 
in the scalar state is stronger than that in the relative spin-1 state.} 
If $\alpha_{\rho N} \neq \alpha_{\lambda N}$, a mixed-symmetry component is generated in the nucleon, but 
existing quark potential models do not support such a contribution of a probability larger than
$1 \div 2 \%$ \cite{fabio}. As can be seen from Eqs. (\ref{nucleon}) and (\ref{3.6}), 
$\alpha_{\rho N}$ is the size parameter of the diquark for which we will use 
the notation $\alpha_D$. 

The wave function (\ref{nucleonwf}) leads to the following expression for the 
proton electromagnetic form factor 
\begin{eqnarray}
\label{protonff}
F_p(Q^2)=\frac23\exp\left(-\frac{Q^2 \alpha^2_{\lambda N}}{12}\right)
+\frac13 \exp\left(
-\frac{Q^2}{48} (\alpha^2_{\lambda N}+3\alpha^2_{\rho N})
\right), \qquad Q^2=\vec q^2.  
\end{eqnarray}
The derivative of the form factor at $Q^2=0$ is experimentally known to be close 
to $1/{M_\rho^2}$, with $M_\rho$ denoting the $\rho$-meson mass. Thus we get a constraint on 
the parameters $\alpha_{\rho N}$ and $\alpha_{\lambda N}$
\begin{eqnarray}
\label{radius}
-F'_p(0)=\frac{\alpha^2_{\lambda N}}{16}+\frac{\alpha^2_{\rho N}}{48}\simeq \frac1 {M_\rho^2}.  
\end{eqnarray}
For the fully symmetric wave function one gets $\alpha_{\rho N}=\alpha_{\lambda N}=2\sqrt{3}/M_\rho=0.9$ fm. 
The nonrelativistic formula (\ref{protonff}) gives a good description of the proton form factor 
in the region $Q^2=0\div 0.3$ GeV$^2$, relevant for our analysis.

\subsection{The pentaquark}

As described in the introduction, we consider the pentaquark as consisting of two 
spin and isospin zero diquarks in a relative $L=1$ angular momentum, and a strange antiquark $\bar s$.  
Thus we take the following form for the $J=1/2$ pentaquark state
\begin{eqnarray}
|\theta^{\uparrow}(\vec p)\rangle&=&\int d\vec r_1 d\vec r_2 d\vec r_3 d\vec r_4 d\vec r_5
\exp\left(i\vec p\;\frac{m_s\vec r_1+m \vec r_2+m\vec r_3+m \vec r_4+m \vec r_3  }{m_s+4m}\right)
\nonumber\\
&&\times\epsilon^{abc}
\left[
\sqrt{\frac{2}{3}}\hat s^{a\downarrow}(\vec r_1) \Psi_{1,1}
-
{\frac{1}{\sqrt3}} \hat s^{a\uparrow}(\vec r_1) \Psi_{1,0}
\right]D^{b}(\vec r_2, \vec r_3) D^{c}(\vec r_4, \vec r_5)
\,|0\rangle. 
\end{eqnarray}
Here we introduced again the creation operator $D^a(\vec r_i, \vec r_j)$ of two correlated quarks. 
The radial wave functions $\Psi_{1,i}$ are angular momentum 1 eigenstates with projection $i$ 
to the $z$-axis. The function $\Psi_{1,i}$ should be symmetric under  
$2\leftrightarrow 3$ and under $4\leftrightarrow 5$ to provide zero values for the diquark spin 
and isospin, and antisymmetric under $\{23\}\leftrightarrow\{45\}$ (i.e. in diquark space coordinates). 
Let us now introduce the variables 
\begin{eqnarray}
\label{3.11}
\vec{r}_{23}=\vec{r}_2-\vec{r}_3, \qquad 
\vec{R}_{23}=\frac{1}{2}(\vec{r}_2+\vec{r}_3), \qquad
\vec{r}_{45}=\vec{r}_4-\vec{r}_5, \qquad
\vec{R}_{45}=\frac{1}{2}(\vec{r}_4+\vec{r}_5), \nonumber\\
\vec{\rho}_\theta=\vec{R}_{23}-\vec{R}_{45}, \qquad \vec\lambda_\theta=\frac{1}{2}(\vec{R}_{23}+\vec{R}_{45})-\vec{r}_1 ~, 
\end{eqnarray}
where $\vec r_1$ is the position of the strange particle, $\vec R_{23}$ and $\vec R_{45}$ are the 
positions of the two diquarks. Taking into account the above symmetry requirements, 
the $P$-wave radial wave function of the pentaquark can be written as follows 
\begin{eqnarray}
\Psi_{1,1}(r_1,r_2,r_3,r_4,r_5)&=&i\frac{\rho_{x+iy}}{\sqrt{2}}\Psi_\theta(r_1|r_2,r_3|r_4,r_5),\nonumber\\ 
\Psi_{1,0}(r_1,r_2,r_3,r_4,r_5)&=&-i\rho_{z}\Psi_\theta(r_1|r_2,r_3|r_4,r_5),  
\end{eqnarray}
where $\Psi_\theta$ is the radial-symmetric part of the pentaquark wave function.  
The diquark picture then amounts to the following form for the latter 
\begin{eqnarray}
\label{psitheta}
\Psi_\theta(r_1|r_2,r_3|r_4,r_5)=
\Phi_\theta(\rho_\theta,\lambda_\theta)\Phi_D(r_{23})\Phi_D(r_{45}),  
\end{eqnarray}
Eq. (\ref{nrnorma}) requires the normalization  
\begin{eqnarray}
\label{pqnorma}
\int |\Phi_D(r)|^2\;d\vec r=1, \qquad
64\int \vec{\rho}_\theta^2 |\Phi_\theta(\rho_\theta,\lambda_\theta)|^2\;d\vec \rho_\theta\, d\vec \lambda_\theta=1, 
\end{eqnarray}
Again, we take a Gaussian parameterizations for the wave function
\begin{eqnarray}
\label{pqwf}
\Psi_\theta(r_1|r_2,r_3|r_4,r_5)=\frac{1}{\sqrt{N_\theta}}
\exp\left(-\frac{1}{2\alpha^2_{\rho \theta}}\vec{\rho}_\theta^2 
-\frac{2}{3\alpha^2_{\lambda \theta}}\vec{\lambda}_\theta^2\right)
\exp\left(-\frac{\vec{r}^2_{23}}{2\alpha_D^2}\right)
\exp\left(-\frac{\vec{r}^2_{45}}{2\alpha_D^2}\right),   
\end{eqnarray}
and determine the normalization factor $N_\theta$ from (\ref{pqnorma}).


\subsection{The transition amplitude}
In the non-relativistic approximation the quark current 
operators take the following form (summation over color indices is implied) 
\begin{eqnarray}
\bar s(\vec x_s)\gamma_0\gamma_5 d(\vec x_d)&\to&
-\hat s^{\uparrow}(\vec x_s)d^{\downarrow}(\vec x_d)+\dots, 
\nonumber\\
\bar s(\vec x_s)\gamma_3\gamma_5 d(\vec x_d)&\to&
\left(\frac{k_x-ik_y}{2m_d}-\frac{k'_x-ik'_y}{2m_s}\right)\hat s^{\downarrow}(\vec x_s)d^{\downarrow}(\vec x_d)
-\left(\frac{k'_z}{2m_s}+\frac{k_z}{2m_d}\right)\hat s^{\uparrow}(\vec x_s)d^{\downarrow}(\vec x_d)+\dots,
\nonumber\\
i\bar s(\vec x_s)\gamma_2\gamma_5 d(\vec x_d)&\to&
\left(\frac{k_x-ik_y}{2m_d}-\frac{k'_x+ik'_y}{2m_s}\right)\hat s^{\uparrow}(\vec x_s)d^{\downarrow}(\vec x_d)
+\left(\frac{k'_z}{2m_s}-\frac{k_z}{2m_d}\right)\hat s^{\downarrow}(\vec x_s)d^{\downarrow}(\vec x_d)+\dots. 
\end{eqnarray}
Here $\vec k'$ is the momentum operator of the $s$-quark,  
$\vec k'=-i\vec \nabla^{(s)} $; 
$\vec k$ is the momentum operator of the $d$-quark, 
$\vec k=-i\vec \nabla^{(d)}$. The dots stand for structures which contain the operators $s^\uparrow d^\uparrow$ or 
$s^\downarrow d^\uparrow$. Their contributions to the $\theta\to N^\uparrow$ 
transition amplitudes vanish: $\langle N^\uparrow|s^\uparrow d^\uparrow|\theta\rangle=0$ and 
$\langle N^\uparrow|s^\downarrow d^\uparrow|\theta\rangle=0$. 

The $\theta\to N^\uparrow$ transition amplitudes of the quark currents above may be expressed through  
\begin{eqnarray}
\vec {\cal A}&=&
\frac{24}{\sqrt{3}}\int d\vec r_2 d\vec r_4 d\vec r_5 
\exp\left(i\vec q\,\frac{\vec r_2+\vec r_4+\vec r_5}{3}\right)\;
\vec \rho_\theta\;\Psi_\theta(r_s|r_2,r_d|r_4,r_5)\left\{2\Psi_N(r_2|r_4,r_5)+\Psi_N(r_4|r_2,r_5)\right\}, 
\end{eqnarray}
with $\vec \rho_\theta=\frac12(\vec r_2+\vec r_d-\vec r_4-\vec r_5)$ as in (\ref{3.11}). The 
functions $\Psi_N$ and $\Psi_\theta$ are given by Eqs. (\ref{psin}) and  (\ref{psitheta}).  
The quantity $\vec {\cal A}$ follows after the sum over color and flavour is performed. 
The two terms in the curly brackets have the following origin: the first (with the factor 2)  
arises from the processes in which one of the diquarks is a spectator.  
The second term is due to the formation of the diquark in the proton by two quarks from 
different diquarks of the pentaquark. 

The final result reads:  
\begin{eqnarray}
|\vec q| F_0=\langle N^\uparrow|\bar s\gamma_0\gamma_5d|\theta^\uparrow\rangle&=&
-i{\cal A}_z|_{\vec r_s\to 0,\; \vec r_d\to 0}
\nonumber
\\
A_L=\langle N^\uparrow|\bar s\gamma_3\gamma_5d|\theta^\uparrow\rangle&=&
 \frac{1}{2m_s}\left\{
 \frac{\partial}{\partial x_s}{\cal A}_x 
+\frac{\partial}{\partial y_s}{\cal A}_y 
+\frac{\partial}{\partial z_s}{\cal A}_z 
\right\} 
\nonumber\\
&+&
\frac{1}{2m_d}\left\{
-\frac{\partial}{\partial x_d}{\cal A}_x
-\frac{\partial}{\partial y_d}{\cal A}_y
+\frac{\partial}{\partial z_d}{\cal A}_z
\right\}|_{\vec r_s\to 0,\; \vec r_d\to 0}
\nonumber
\\
\label{3.20}
A_T=i\langle N^\uparrow|\bar s\,\gamma_2\gamma_5d|\theta^\downarrow\rangle&=&
\frac{1}{2m_s}\left\{
 \frac{\partial}{\partial x_s}{\cal A}_x
+\frac{\partial}{\partial y_s}{\cal A}_y
+\frac{\partial}{\partial z_s}{\cal A}_z
\right\}
\nonumber\\
&+&
\frac{1}{2m_d}\left\{
-\frac{\partial}{\partial x_d}{\cal A}_x
+\frac{\partial}{\partial y_d}{\cal A}_y
-\frac{\partial}{\partial z_d}{\cal A}_z
\right\}|_{\vec r_s\to 0,\; \vec r_d\to 0}. 
\end{eqnarray}
\section{Numerical estimates}
We can give now our numerical results for the pentaquark width using (\ref{ga}) and (\ref{3.20}). 
Two assumptions reduce the number of parameters:   

1. The structure of the diquark in the nucleon and in the pentaquark coincide, i.e. we take 
the size-parameter $\alpha_D$ of the diquark wave function $\Phi_D$ to be equal to the parameter 
$\alpha_{\rho N}$ of the nucleon wave function 
\begin{eqnarray}
\alpha_D=\alpha_{\rho N}. 
\end{eqnarray}

2. The parameters of the nucleon wave function are chosen such that the experimental 
nucleon electromagnetic form factor is reproduced for small momentum transfers,   
\begin{eqnarray}
\label{nucleonparam}
\frac{\alpha^2_{\lambda N}}{16}+\frac{\alpha^2_D}{48} = \frac1{M_\rho^2}.  
\end{eqnarray}
We first take a symmetric wave function $\alpha_{\lambda N}=\alpha_D$.
As explained above, the diquark size parameter is then given by the relation 
$\alpha_D=\alpha_{\rho N}=0.9$ fm, a number also compatible with 
the size of the pion. Now only the two free parameters of the pentaquark wave function 
remain to be fixed, namely, $\alpha_{\theta \rho}$ and $\alpha_{\theta \lambda}$. 
We study the dependence of  
$g_A$ and of the width on these parameters. We recall that the average distance between the 
two diquarks is determined by $\alpha_{\theta\rho}$, and the average distance between 
the $s$-antiquark and the center-of-mass of the two diquarks is determined by $\alpha_{\theta\lambda}$.

The decay width involves the constituent quark masses $m_s$ and $m_u=m_d$. 
We use the values $m_s\simeq 500$ MeV and $m_d\simeq 350$ MeV 
relevant for a nonrelativistic description. We have checked that 
the values $m_s\simeq 350$ MeV and $m_d\simeq 220$ obtained in a relativistic quark model 
\cite{m} essentially lead to the same results, since the $d$ 
and $s$ quark contributions to the amplitudes partially compensate each other. 

Little is known about the details of the pentaquark structure. We allow therefore the parameters 
$\alpha_{\theta\lambda}$ and $\alpha_{\theta\rho}$ to vary in a broad range  
\begin{eqnarray}
0.6\; fm<\alpha_{\theta\lambda},\;\alpha_{\theta\rho} < 1.6\; fm. 
\end{eqnarray}
If both parameters would be 1 fm, then $g_A\simeq 0.8$ and the corresponding width is $150$ MeV. 
No suppression due to a possible mismatch of color and flavour quantum numbers in the 
initial and final states takes place. 

However, a strong dynamical suppression occurs if the structure of the pentaquark is asymmetric. 
Figure 1 exhibits the dependence of $g_A$ on the pentaquark size parameters. For instance, for 
\begin{eqnarray}
\label{ppp} 
\alpha_{\theta\lambda}=0.6 \; fm,\qquad \alpha_{\rho\lambda}=1.4 \; fm, 
\end{eqnarray}
we get $g_A=0.05$ and $\Gamma(\theta)=1$ MeV. 
We note that a somewhat small $\alpha_{\theta\lambda}$ and a large $\alpha_{\theta\rho}$
imply an interesting "peanut" structure of the pentaquark: 
the two extended diquark balls overlap only partially, while the antiquark remains close to the center.

Figure 2 presents the form factor $g_A(q^2)$ for these values of the parameters. We observe that 
the form factor is a rather flat function of $q^2$ as expected from the location of 
the $K^*_A$ resonance. 

It is also of interest to see the influence of an extreme $qq$ correlation (i.e. of a very small diquark size) 
on the pentaquark decay width.  
Figure 3 illustrates the pentaquark width vs the diquark size $\alpha_D$ 
for fixed values of the pentaquark size-parameters. It is seen that 
a sizeable reduction of the pentaquark width occurs only for a very small diquark size,
which corresponds to an implausible large deviation from an SU(6)-symmetric nucleon wave 
function. 
The successful description of the nucleon using a symmetric nucleon wave function does 
not support such small-size diquarks. The diquark should be an extended object which 
dissolves into smaller partons in higher momentum transfer processes. 
\begin{figure}[ht]
\begin{center}
\begin{tabular}{cc}
\includegraphics[totalheight=5cm]{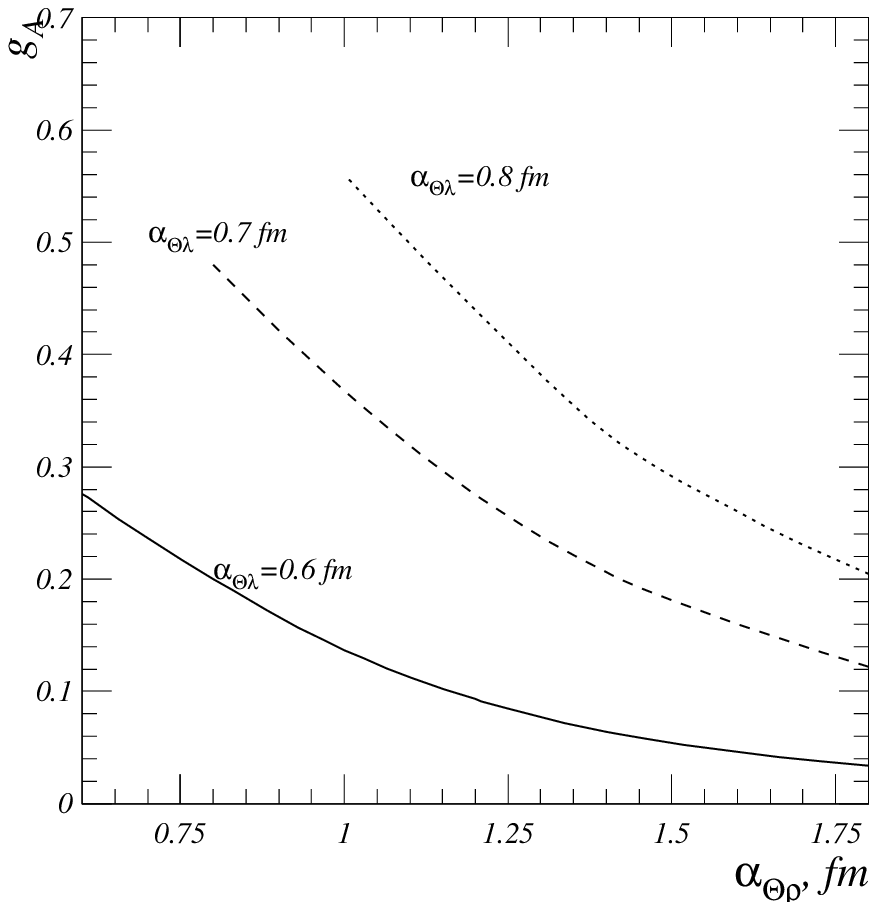} 
&
\hspace{1cm}\includegraphics[totalheight=5cm]{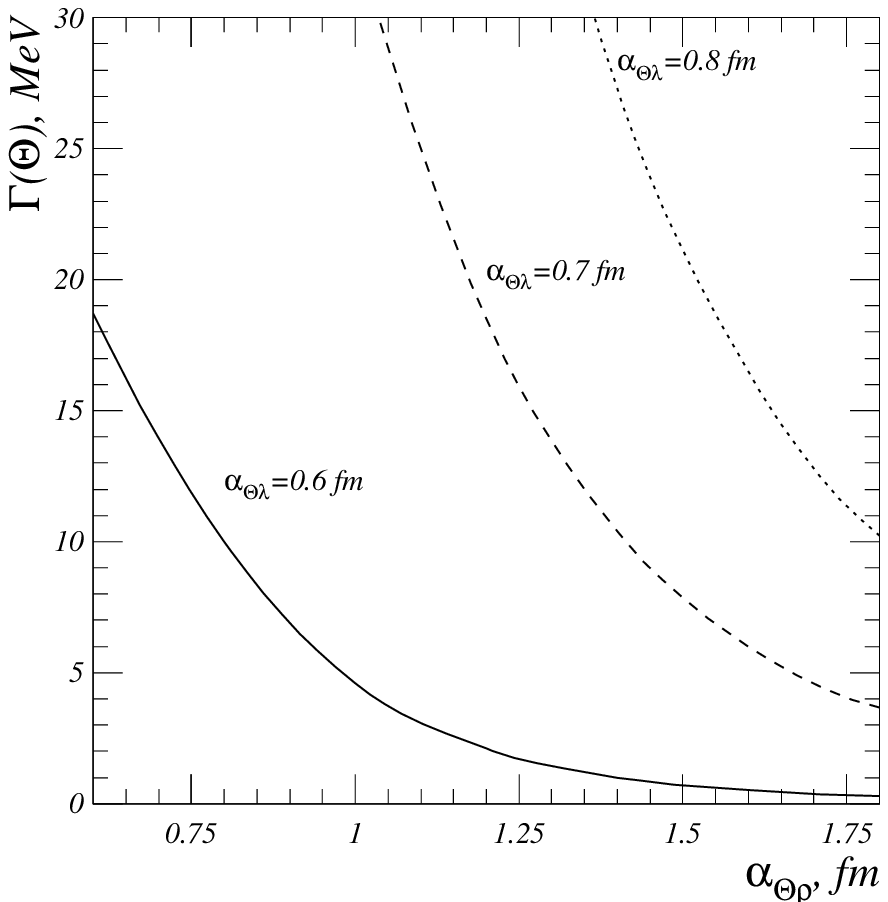}\\
\end{tabular}
\caption{\label{fig:1}
The axial-vector coupling $g_A$ and the decay width   
$\Gamma(\theta)$ vs 
the pentaquark size parameter $\alpha_{\theta\rho}$ for different values of $\alpha_{\theta\lambda}$.} 
\end{center}
\end{figure}
\begin{figure}[ht]
\begin{center}
\begin{tabular}{c}
\includegraphics[totalheight=5cm]{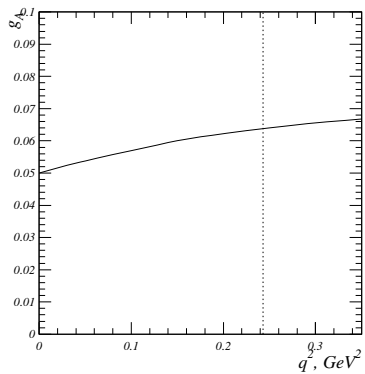} 
\end{tabular}
\caption{\label{fig:2}
The axial-vector form factor $g_A$ vs $q^2$ for the pentaquark size parameters 
$\alpha_{\theta\rho}=1.4$ fm and $\alpha_{\theta\lambda}=0.6$ fm. The dotted line corresponds to $q^2=M_K^2$.} 
\end{center}
\end{figure}
\begin{figure}[ht]
\begin{center}
\begin{tabular}{c}
\includegraphics[totalheight=5cm]{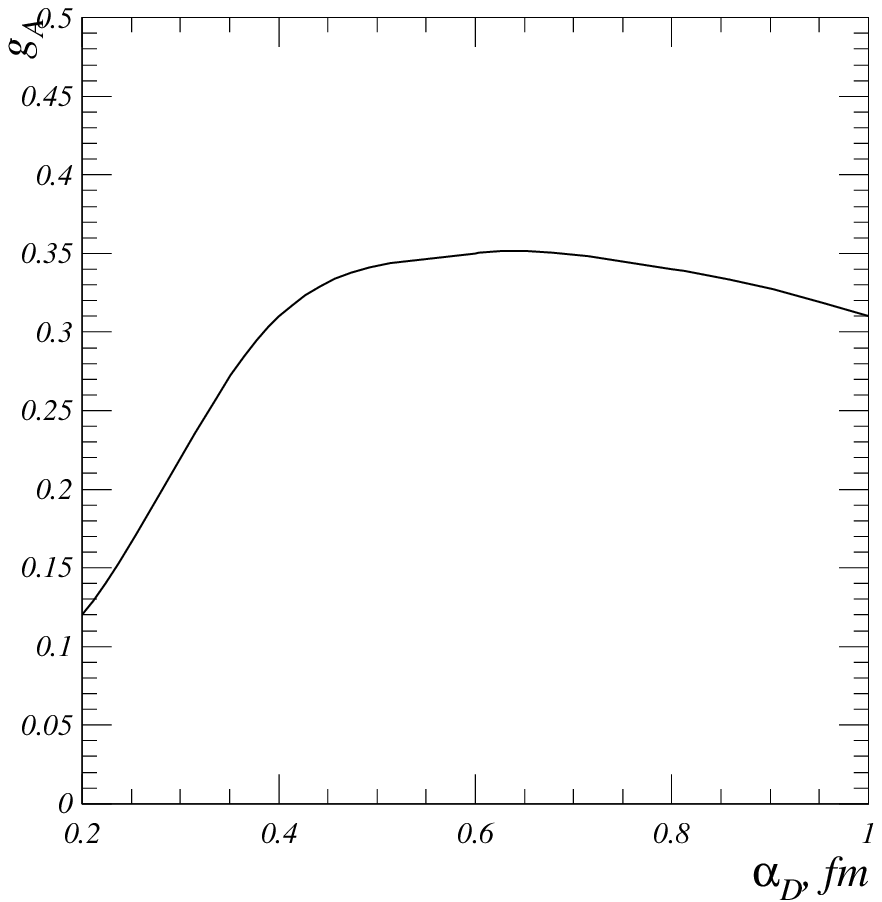} 
\end{tabular}
\caption{\label{fig:3}
The axial-vector coupling $g_A$ vs the diquark size parameter $\alpha_D=\alpha_{N\rho}$. 
The parameter $\alpha_{N\lambda}$ of the nucleon wave function is calculated from 
$\alpha_{N\rho}$ using Eq. (\ref{radius}). 
The values of the pentaquark size parameters are fixed to be 
$\alpha_{\theta\rho}=1.2$ fm and $\alpha_{\theta\lambda}=1.0$ fm.} 
\end{center}
\end{figure}
\section{Conclusions}
We studied the $\theta({\frac12}^+)$ pentaquark decay into $NK$ final states,  
assuming that the pentaquark consists of two extended color-triplet $ud$ diquarks with spin and isospin zero 
in a relative $P$-wave state, and a strange antiquark. 
The pentaquark and the nucleon were described in terms of non-relativistic Fock states with Gaussian 
spatial wave functions. 

The pentaquark decay width $\Gamma(\theta)$ is found to depend strongly on the pentaquark configuration: 
when all size-parameters of the pentaquark wave function are close to 1 fm, one obtains a width of 
about $100 - 150$ MeV, i.e. a typical hadronic value. The color-flavour structure of the pentaquark causes no suppression. 

However, a strong dynamical suppression of the amplitude occurs for a "peanut"-shaped  
pentaquark, i.e. when it has an asymmetric structure with 
$\alpha_{\theta \lambda}\ll \alpha_{\theta \rho}$. 
For instance, $\alpha_{\theta \lambda}=0.6$ fm and $\alpha_{\theta \rho}=1.4$ fm 
brings the width down to 1 MeV. 

Because of the specific details of our analysis (non-relativistic treatment, 
Gaussian wave functions) the accuracy of our calculation is presumably only about 50\%. 
Nevertheless, the proposed mechanisms for a strong suppression of the amplitude 
by a specific peanut-like structure of the pentaquark remains valid. 

We conclude therefore that if the pentaquark can be described as a quark-diquark system, 
the small width requires a rather asymmetric structure with two 
extended diquarks rotating about the strange antiquark near the center. 


{\it Acknowledgments.} We thank Dieter Gromes and Otto Nachtmann for useful discussions. 
The work was supported by Bundesministerium f\"ur Bildung und Forschung 
(BMBF) under project 05 HT 1VHA/0. 


\end{document}